\documentclass{aa}
\input epsf.sty
\usepackage{graphics}
\usepackage{times}

\newcommand{\lsi}    {LS~I+61$^{\circ}$303}
\newcommand{\lsa}    {LS~I+61$^{\circ}$235}
\newcommand{\Ha}     {H$\alpha$}
\newcommand{\vsi}    {$\;v \sin{i}$} 
\newcommand{\kms}     {~km~s$^{-1}$}

\newcommand{\gtsima} {$\; \buildrel > \over \sim \;$}
\newcommand{\simgt}  {\lower.5ex\hbox{\gtsima}}            
\def\simless{\mathbin{\lower 3pt\hbox
     {$\rlap{\raise 5pt\hbox{$\char'074$}}\mathchar"7218$}}}   
\def\simmore{\mathbin{\lower 3pt\hbox
     {$\rlap{\raise 5pt\hbox{$\char'076$}}\mathchar"7218$}}}   

\begin{document}

\thesaurus{06     
        ( 08.05.2      
          08.03.4        
          08.02.1        
          13.25.5        
        ) }

  \title{ Comparison of the  H$\alpha$ circumstellar disks in Be/X-ray binaries 
           and Be stars }

  \author {  R.K.~Zamanov\inst{1}
            \and  P.~Reig\inst{2,3}
            \and  J.~Mart\'{\i}\inst{4}
            \and  M.J.~Coe\inst{5} 
            \and J.~Fabregat\inst{6}
            \and N.A.~Tomov\inst{1}
            \and T.~Valchev\inst{7}
           }
         
  \offprints { R.Zamanov,  rozhen@mbox.digsys.bg }

   \institute{ Institute of Astronomy, Bulgarian Academy of Sciences
 and Isaac Newton Institute of Chile, Bulgarian Branch, 
National Astronomical Observatory Rozhen, 
P.O.Box 136, BG-4700 Smolyan, Bulgaria (e-mail: rozhen@mbox.digsys.bg)  
         \and  Foundation for Research and Technology-Hellas, 711~10 Heraklion, Crete, Greece
         \and  Physics Department, University of Crete, 710~03 Heraklion, Crete, Greece
         \and  Departamento de F\'{\i}sica, EPS,
               Universidad de Ja\'en, C/ Virgen de la Cabeza, 2, E--23071 Ja\'en, Spain
               (jmarti@ujaen.es)
         \and  Physics and Astronomy Department, Southampton University, 
               Southampton, S017~1BJ, UK
         \and  Departamento de Astronom\'{\i}a, Universidad de Valencia,
               E--46100 Burjassot, Valencia, Spain
         \and National Astronomical Observatory Rozhen, 
              P.O.Box 136, BG-4700 Smolyan, Bulgaria     }

\date{Received 2 June 2000 / Accepted 15 December 2000 } 

\maketitle

\markboth{Zamanov et al. :  \Ha\ cicumstellar disks in Be/X-ray binaries }{}

\begin{abstract} 

We present a comparative study of the circumstellar disks in Be/X-ray
binaries  and isolated Be stars based upon the \Ha\ emission line.
From this comparison it follows that the overall structure of the
disks in the Be/X-ray binaries is similar to the disks of other Be stars,
i.e. they are axisymmetric and rotationally supported. 
The factors for the line broadening  (rotation and temperature)  in the disks
of the Be stars and the Be/X-ray binaries  seem to be identical.
 
However, we do detect some intriguing differences between the envelopes.
On average, the circumstellar disks of the Be/X-ray binaries are twice as dense
as the disks of the isolated Be stars.  The different distribution of
the Be/X-ray binaries and the Be stars seen in the full width half
maximum versus peak separation diagram 
indicates that the disks
in Be/X-ray  binaries have on average a smaller size, probably truncated 
by the compact object.\\


\end{abstract}

\section{Introduction}
Be stars are early type non-supergiant stars 
surrounded by a circumstellar envelope with disk-like geometry.
The circumstellar disk is visible in the Balmer 
(and sometimes other) line emission and also gives 
rise to a strong infrared 
excess. The physical mechanism of the disk formation is not well 
understood, but it is generally believed that the fast rotation 
of the central star plays an important role. 
The Be phenomenon is present in both isolated B type stars 
and X-ray binaries with massive B type companions. 
The Be/X-ray binaries represent  the major subclass of massive X-ray 
binaries, in which the X-ray emission is due to the interaction of
the compact object (usually a neutron star) with the wind of the Be star.

In this paper, we explore the existence of significative differences
between the Be disks in both kinds of systems. The black hole, or neutron
star, companion in an X-ray binary may be affecting the Be disk structure.
Is there any observational evidence? Can we see the signature of the
compact companion in the Be emission line profiles?  Traditionally, the
answer to this question has been negative, and it was believed that the
presence of the compact object would not affect the dynamics of the Be
disk.  However, recent results seem to contradict this idea.  Reig,
Fabregat \& Coe (1997) found a relationship between the properties of the
disk and the orbit of the neutron star. They reported that Be stars 
in X-ray binary systems have, on average, a lower \Ha\ equivalent width
(EW) when compared to a complete set -- those contained in the Bright Star
Catalogue (Hoffleit \& Jascheck 1982) -- of isolated Be stars. They also
discovered a possible correlation between the maximum observed EW(\Ha) and
the orbital period of the system, which was explained by assuming that the
compact object truncates the outer parts of the Be star's disk. The idea
of disk truncation, together with disk warping has been used by Negueruela
et al. (2000) to explain the X-ray outburst mechanism in the Be/X-ray
system \object{4U~0115+63} (\object{V635 Cas}).

The solution to these questions is likely to be provided
by a careful and exhaustive examination of high dispersion
optical and infrared (IR) spectra at high signal-to-noise ratio (S/N). 
The following sections are devoted to a comparative 
study of the circumstellar disks in the "normal" Be stars 
and the Be/X-ray binaries based on the best available 
\Ha\ emission line spectroscopy.

\section{ Spectroscopic Data }

In our comparative study we used the \Ha\ equivalent width, full width
at half maximum (FWHM) and the distance between the peaks 
($\Delta$V). The data for the `normal' Be stars were taken from the high
resolution,  high S/N observations of  
Andrillat (1983), Hanuschik (1986), Hanuschik, Kozok \& Kaiser (1988),
Dachs, Hummel \& Hanuschik (1992),  Slettebak, Collins \& Truax (1992).  The
optical counterparts of the Be/X-ray binaries have spectral classes in the
range O8-B2, and represent a significantly different 
distribution from that of
isolated Be stars (Negueruela 1998). But as far as we know there is no
difference between the disk structure of the Be stars with different
spectral types, so we will use all  the data from the papers referred
to above. 

The H$\alpha$ parameters corresponding to the Be/X-ray binaries were taken
from  the  Southampton/Valencia/SAAO data base of optical and IR
observations (Reig et al. 1997). 
This data base covers more than 10 years of
observations  from a variety of telescopes in both the northern and
southern hemispheres.  Additional observations were made from the
Bulgarian National Astronomical Observatory `Rozhen' with the Coud\'e
spectrograph of the  2.0m RCC telescope giving a  
dispersion of 0.2 \AA\ pix$^{-1}$. 
The results of our measurements are summarized in Tables 1 and
2. Our intention was to measure the EW(H$\alpha$) as defined by 
Dachs et al. (1981). Because traces of residual wings 
due to photospheric 
absorption were not detected, the interpolated continuum was used as 
baseline during the measurements. 

In the case of double peak
profiles the parameter FWHM is measured only when the both peaks have
an intensity 
above the half-maximum. This is because the parameter is 
used to investigate the broadening in the disk and 
it would give unusual values if measured using only one peak, 
i.e. it would be sensitive to the broadening process 
from half of the disk only. The last column in the Tables 1 and 2 represents
the reciprocal dispersion of the spectra in \AA\ per pixel. 

Additionally to these observations we used the data for 
\object{4U~1145-619}  from Cook \& Warwick (1987), 
\object{LS~I+61$^{0}$303} by Zamanov et al. (1999),
and \object{LS~I+61$^0$235} by Reig et al. (2000a). 
The values used for the projected
rotational velocity, \vsi,  are summarized in Table~3.

\section{ Comparison of the $H\alpha$ line parameters for
Be and Be/X-ray stars }

In this section we present the behaviour of the \Ha\ parameters, one
versus another, for isolated Be stars and Be/X-ray binaries. 

In an attempt to
provide a quantitative approach to our study, we have performed two
dimensional Kolmogorov-Smirnov  tests (Peacock 1983, Press et al. 1994) on
every figure. This test gives us the probability that both samples are
compatible with each other. Usually if the indicated probability
is  $< 0.05$ the two data sets are statistically different. 
The Be stars atlases contain a lot of stars,
each one observed one or several times. However for Be/X-ray binaries,  we
have merely 10 objects observed a lot of times. Therefore in the KS-test we used
only  2-4 points for every Be/X-ray binary, in order to have 
a similar sample from each group. Thus the
effective number of the data points is \simgt 20,
which should give sufficient accuracy for the test.

It deserves to be noted that in general 
the type of the profile (double or single peak)  and 
the V/R ratio can influence the positions in the diagrams, 
e.g. Dachs et al. (1992) detected a relationship between 
the V/R ratio and the radial velocities of the peaks.
Our attempts to use only single peak profiles
or only almost symmetrical profiles (V/R $\sim$ 1.0$\pm$0.2) do not
give different pictures. So in the following we will use all 
data independently of the \Ha\ profile type .

  \begin{figure}[htbp]
     \normalsize
     \vskip 3mm plus 1mm minus 1mm
           \epsfysize=14.5cm
           \centerline{\epsffile{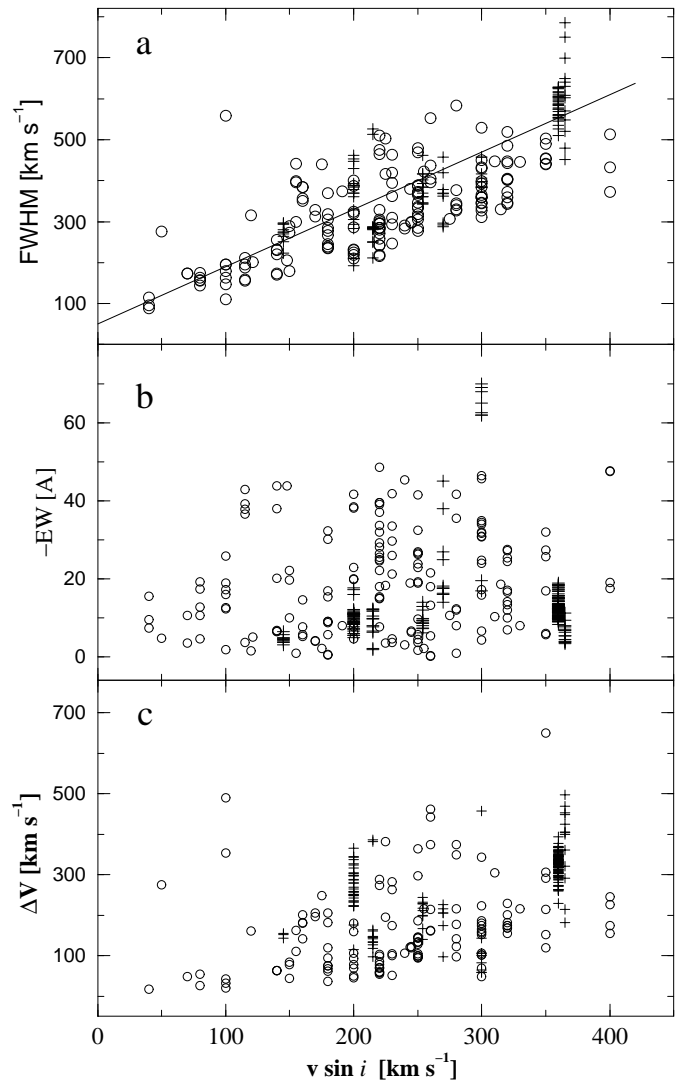}}
  \caption[] { {\bf a-c.} The \Ha\ line parameters versus \vsi. 
The circles represent the Be stars, the crosses - the Be/X-ray binaries.

{\bf a } FWHM versus \vsi.
The solid line is \\ 
$FWHM=1.4\:v\:\sin\:i\;+\;50\:$km s$^{-1}$, 
which is the average behaviour of the Be stars, as defined by 
Hanuschik (1989). 

{\bf b } EW(\Ha) versus \vsi.

{\bf c } The distance between the peaks versus \vsi.
    }    
  \label{Xvsini}
  \end{figure}

\subsection{ The \Ha\ parameters versus \vsi }

Fig.~\ref{Xvsini}$a$ shows the behaviour of the \Ha\ full width at half
maximum (FWHM). It is well known that the width of  the \Ha\ emission line
increases linearly with   $\;v \sin{i}\;$ (Struve 1931, Andrillat \&
Fehrenbach 1982, Dachs et al., 1986). 

The observed points for Be/X-ray binaries are well inside the scatter 
of other  Be stars. The KS test gives a significant probability of 0.10-0.15, 
that the points are extracted from the same distribution. 

The observed correlation between stellar rotational velocities, \vsi, of
the Be stars and FWHM of their emission-line profiles gives the strongest
evidence for a rotationally supported disk-like circumstellar envelope
and a coupling of photospheric and envelope rotation (e.g. Dachs et al. 1986; 
Hanuschik, 1989). The fact that the Be/X-ray binaries have a similar 
distribution in Fig.~\ref{Xvsini}$a$ points that, independently
of the presence of compact companion, their envelopes are 
axisymmetric and rotationally supported, as could be expected.

In Fig.\ref{Xvsini}$b$ and Fig.\ref{Xvsini}$c$, we plot the behaviour of
the EW and  the peak separation, $\Delta$V, versus \vsi, respectively.  In
Fig.1b the KS test gives a probability of 0.01-0.001, which suggests
strongly that both samples are statistically different. 
Our numerical experiments suggest that this is mainly due to 
most Be/X-ray binaries not achieving EW(\Ha) values 
above 25 \AA. Some influence on the KS-test results has also 
the fact that low values of \vsi\ are missing among 
the observed Be/X-ray binaries, but this has minor contribution. 
Therefore, Fig.1b supports the
original result by  Reig, Fabregat \& Coe (1997) that the Be/X-ray
binaries have, on average, a lower EW(\Ha) when compared to a
set of isolated Be stars. 

In Fig.~\ref{Xvsini}$c$ the KS test gives a probability of 0.2-0.3, which
indicates that the behaviour of the peak separation for  both types is
very similar. It  is worth noting that (although it is not statistically
significant), the peak separation of  the Be/X-ray binaries does not
achieve values  smaller than the peak separation of the Be stars
($\Delta$V is a measure of the disk size, see Eq.\ref{Huang}). 


\subsection{ The \Ha\ width parameters versus the equivalent width }

  \begin{figure}[htbp]
     \normalsize
     \vskip 3mm plus 1mm minus 1mm
           \epsfysize=10.0cm
           \centerline{\epsffile{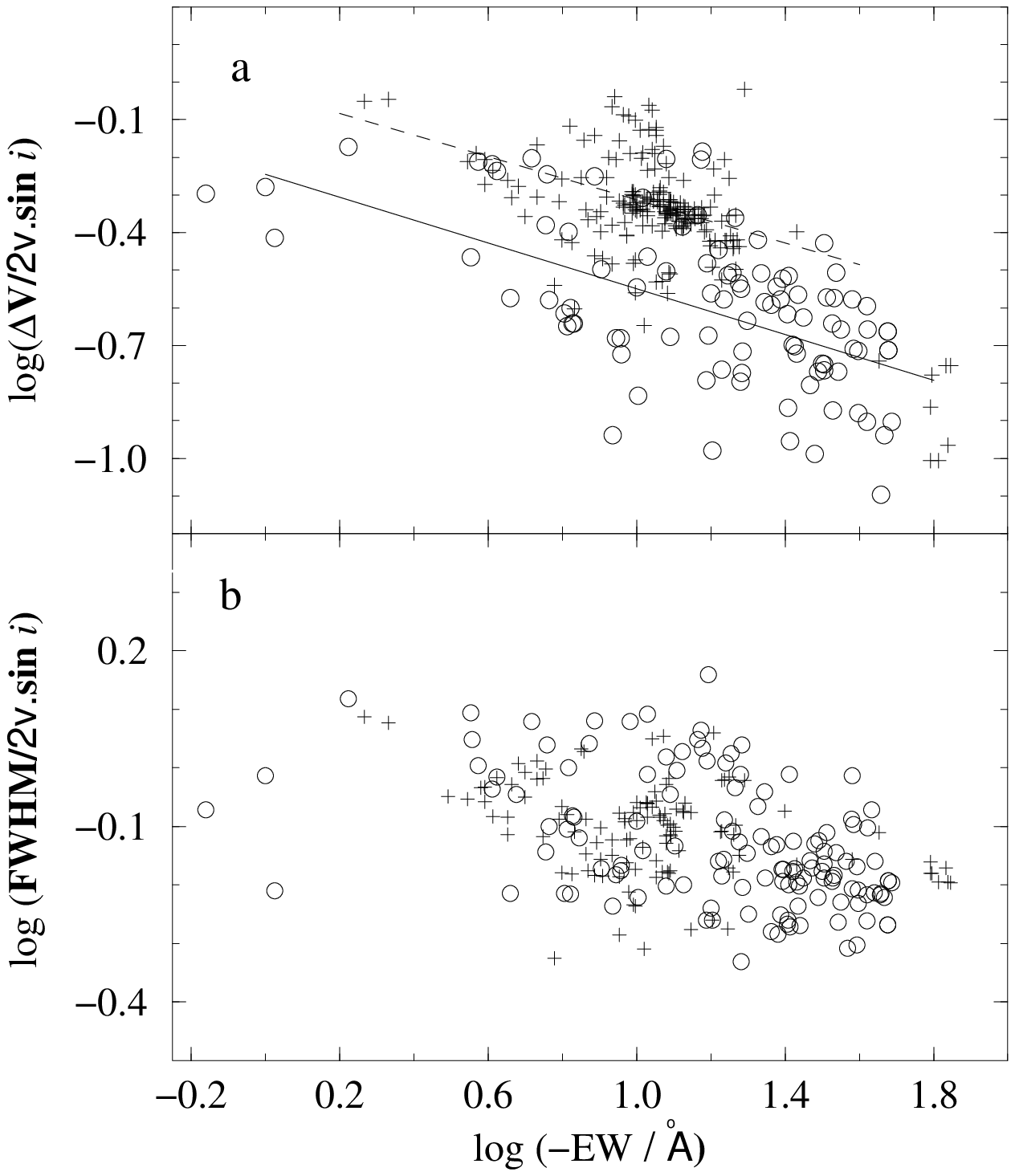}}
  \caption[] {
 {\bf a and b.} The \Ha\ line parameters versus EW(H$\alpha$).  

  {\bf a } Plot of $\log{(\Delta V/(2\:v\sin{i}))}$ versus 
  EW(H$\alpha$). The lines represent the best linear fits:
  the solid line over the circles (Be stars), 
  the dashed line over the crosses (Be/X-ray stars). 
  The best fit line of the Be/X-ray binaries is shifted to 
  denser circumstellar disks.

{\bf b }  $\log{(FWHM/(2\:v\sin{i}))}$ versus $\log{EW(H\alpha)}$. 
  The Be/X-ray binaries and isolated Be stars mixed together.
}    
  \label{XlgEW2}
  \end{figure}

In Fig.\ref{XlgEW2} the \Ha\ width parameters $\Delta V$  and FWHM,
normalized to \vsi,  have been plotted as a function of the EW(\Ha). 
In the top panel, it
is clear that the Be/X-ray data points as well as their linear fit  are
considerably shifted relative to the behaviour of the Be stars.  The KS
test gives a probability of $\sim$0.001 if we use all data points, and
probability $\sim$0.01 if we use only data with EW$\leq$20 \AA.  This fact
can be interpreted as indicating the 
presence of denser disks and will be discussed in
Sect.4.

In the bottom panel, the KS test for EW$\leq$20 \AA\ gives a probability of
0.20-0.40. It is also obvious that both samples occupy one and the same
place at low EW. This result is not unexpected and implies that the
processes relevant to the line broadening (Kepler rotation,  
thermal broadening, non-coherent scattering) 
are identical in the disks of the Be and Be/X-ray stars, when
the latter are non perturbed. 
There are several cases when one of the wings is strongly affected,
and, as a result,
it has an intensity below the half maximum. 
Probably  this is an effect of global one-armed
oscillations (see Negueruela et al. 1998, for a discussion of this
effect). In such cases we did not measure the FWHM
for the reasons mentioned in Sect.2.

\subsection{FWHM versus  $\Delta V$  }

  \begin{figure}[htbp]
     \normalsize
     \vskip 3mm plus 1mm minus 1mm
           \epsfysize=5.8cm
           \centerline{\epsffile{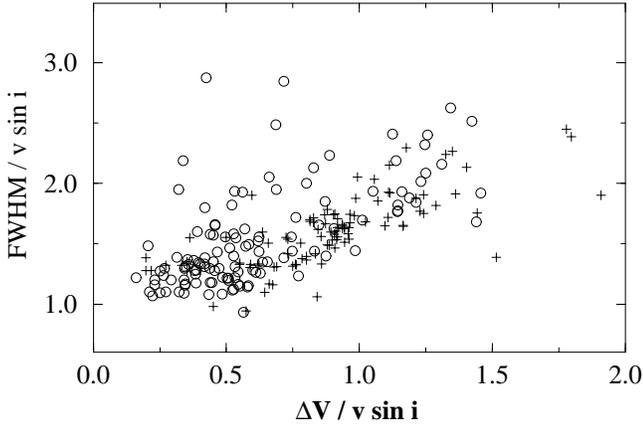}}
  \caption[] { FWHM versus the peak separation, 
  $\Delta V$, normalized with \vsi. 
  The symbols are the same as in Fig.~\ref{Xvsini}.       }    
  \label{FWHMdV}
  \end{figure}

In Fig.\ref{FWHMdV} we plot the FWHM against the peak separation, both
normalized to \vsi. The KS test gives here a probability of about 0.001.
This strongly indicates that the two samples are drawn
from different distributions. 

The FWHM of the \Ha\ emission-line  profiles are essentially determined
by  the effective rotational velocity of those regions of the envelope
which are emitting an important fraction of the \Ha\
radiation.  In Fig.\ref{FWHMdV} the crosses of the Be/X-ray binaries are
shifted towards high $\Delta$V. This implies that, at the same effective
emitting size, their disks are smaller. Such a result, can be interpreted
as a clear indication 
that in many cases  the outer parts of the Be/X-ray disks are missing,
probably truncated by the neutron star. 

\section{ Dense circumstellar disks in the Be/X-ray binaries}

  \begin{figure}[htbp]
     \normalsize
     \vskip 3mm plus 1mm minus 1mm
           \epsfysize=6.2cm
           \centerline{\epsffile{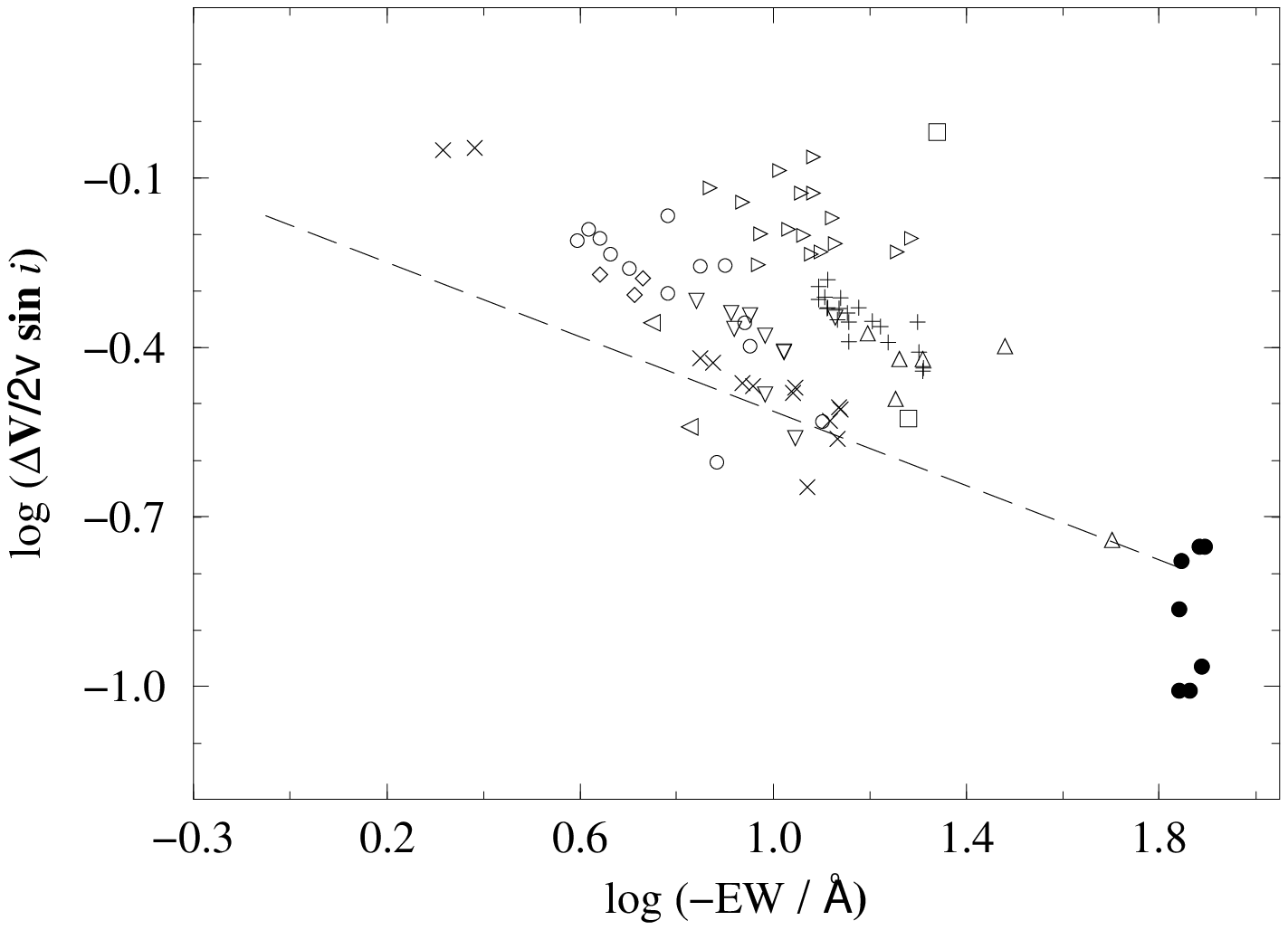}}
    \caption[]{Plot of $\;\log{(\Delta V/2v\:\sin{i})}$ versus 
 $EW($H$\alpha)$ of Be/X-ray binaries. The dashed line
represents the average behaviour of the Be stars
(from Hanuschik, 1989). 
It is apparent that the data points for the Be/X-ray binaries 
are shifted up,  towards denser Be disks. \\
\hskip 0.5cm  {\bf Legend:}  pluses (+) - \lsi,  
crosses ($\times$) - \object{X~Per}, diamonds - \object{BQ Cam}, 
triangles up - 4U~1145-619, triangles down - \object{A~0535+26}, 
triangles right - \lsa, triangles left - \object{4U~0728-25}, 
open circles - V635~Cas, filled circles - \object{A~1118-617}, 
squares - \object{LS~992}.
  }
  \label{XlgEW1}
  \end{figure}

It is known that the \Ha\ peak separation and the equivalent 
width correlate for the Be stars (Hanuschik, Kozok \& Kaiser, 1988). 
Hanuschik (1989) have derived the law
  \begin{equation}
      \log \left( \frac{\Delta V}
                {2\,v\,\sin{i}} \;\right)
       = \;-a\; \log \left( \frac {-EW(H\alpha)}{\AA}\;\right) +\;b ,
  \label{EW-peaks}
  \end{equation}
where the average values are $a \simeq 0.32$ and $b
\simeq -0.2$. It is worth noting that when using almost twice as many points 
we obtained practically the same values $a = 0.34 \pm 0.06$  
and $b=-0.18 \pm 0.07 $.
The law in Eq.~\ref{EW-peaks} expresses that the outer radius 
grows as the $EW($H$\alpha)$ becomes larger, because for rotationally 
dominated profiles the peak separation can be regarded as a measure of 
the outer radius of the \Ha\ emitting disk (Huang, 1972): 
 \begin{equation}
      \left( \frac{\Delta V}
                {2\,v\,\sin{i}} \;\right)
       = \;  \left( \frac {R_{disk}}{R_*}\;\right)^{-j} ,
  \label{Huang}
  \end{equation}
with $j=0.5$ for Keplerian rotation and $j=1$ for conservation 
of angular momentum.

The same quantities for the Be/X-ray binaries are plotted in 
Fig.~\ref{XlgEW1}. The dashed line represent the average behaviour of
Be stars as defined by Hanuschik (1989).
Hanuschik, Kozok \& Kaiser (1988)  consider that the  different
electron density ($N_e$) is the factor responsible for the vertical
scattering in this figure. They estimated that the  limits for the Be
stars correspond to  

\begin{equation} N_{e,max}/N_{e,min}\:\simeq\:4.5. 
\label{Ne} 
\end{equation}

From Fig.~\ref{XlgEW1} it can be seen that almost all the data points 
for the Be/X-ray
binaries are above the average line, i.e. they are  shifted towards the
denser disks.  This indicates  that the circumstellar disks in the
Be/X-ray systems are about $\simeq 2$ times more dense than disks in 
isolated Be stars. 

This result is very unlikely
to be due to the lower resolution of the observations involving Be/X-ray
binaries, because at least for three systems, X~Per, \lsi\ and \lsa, 
some observations had high spectral resolution 
( 0.10 -- 0.067 \AA\ pix$^{-1}$ ) and  
these data do not show significant  differences with  
respect to the lower resolution observations.

This behaviour was
first discovered for \lsi\ (Zamanov et al. 1999) and  for \lsa\ (Reig et
al. 2000a).  Given the large amount of data collected for the systems 
\lsi\ and \lsa\ we randomly selected a subsample  of their observations 
in order not to give too much weight to them in Fig.\ref{XlgEW1}.

The possibility that the disks in the Be/X-ray binaries
are denser has already been 
suggested by Negueruela et al. (1998) in connection 
with the shorter periods of V/R variability.

\section{Discussion}

The Balmer emission lines that characterize the spectra of Be stars
provide information about the physical conditions of the circumstellar
equatorial disk. In this work we have used  the H$\alpha$ emission line
spectral parameters to compare the properties of the disks in isolated Be
stars and Be stars in Be/X-ray binaries. Our goal is to assess the possible
effect of the compact object on the Be star's disk.

In our comparative study, we find similarities but also interesting
differences in the physical properties of both type of systems.

\subsection{Similarities}

The similarities 
seem to be related to the overall structure of the disks, as could 
be expected. Firstly, the disks  in Be/X-ray binaries,
like those in isolated Be stars, 
are axisymmetric and rotationally supported. 
Secondly, isolated Be
stars and Be stars in binaries populate the same place in the FWHM--EW
diagram for low values of the EW.
In other words, the line broadening processes (such as Keplerian rotation,
thermal broadening, non-coherent scattering) are similar. 
Another similarity was presented by Telting (2000). He pointed
out that the disk growth rates 
after a disk-loss episode for stars of similar spectral type are similar:
4$\times$10$^{-9}$~M$_{\odot}$~yr$^{-1}$ for the Be/X-ray binary 
X Per (Telting et al. 1998) and 
3.8$\times$10$^{-9}$~M$_{\odot}$~yr$^{-1}$ for $\mu$ Cen (Hanuschik et al. 1993).

\subsection{Differences}

The differences begin to emerge when the disk develops. 
The results of Reig, Fabregat \& Coe (1997) as well as the KS-test performed 
on Fig.~\ref{Xvsini}$b$ point to that the Be/X-ray binaries show smaller 
EW than isolated Be stars. Only the systems 4U~1145-619 and A~1118-617 
(located at
\vsi=270\kms on Fig.\ref{Xvsini}) and A1118-617 (at \vsi=300 \kms) 
show EW comparable to those
of isolated Be stars. The orbital period of 4U~1145-619 is 188 d and that of 
A~1118-617 is unknown but likely to be of the order of hundreds of days
according to Corbet's diagram (Corbet 1986). With such wide orbits we
would not expect the neutron star to have a strong impact on the disk
evolution. 
In relation with this result there is the possibility that the peak
separation $\Delta$V in Be/X-ray systems tends to be larger than in Be
stars (Fig~\ref{Xvsini}$c$). Although this finding could be due, in
general, to the lower resolution data for Be/X-ray binaries, if real,
it can represent further evidence supporting the idea 
that the Be stars in
X-ray binaries cannot develop extended disks because of its interaction
with the neutron star (provided that $\Delta$V can be considered as a
measure of the disk radius).

The strongest evidence for a different behaviour of the circumstellar
disks in Be and Be/X-ray systems is provided by Fig.~\ref{XlgEW2}$a$ and
Fig.~\ref{XlgEW1}. In the $\Delta$V--EW diagram, Be/X-ray binaries 
are located in a
higher region than isolated Be stars, indicating that they are, on
average, about two times denser.   
Negueruela et al. (1998) noticed that the quasi periods of the V/R 
variability  observed in
Be/X-ray binaries are, on average, much shorter than those  of isolated Be
stars.  They suppose that this difference may be caused by denser envelopes
in the binary systems. Our results seem to confirm such idea. 
Given the tendency of Be/X-ray binaries to occupy a preferable
position in this diagram it would be very interesting if we were able to 
find a correlation between the position of the star in this diagram and
other parameters, such as the Balmer decrement, the period of V/R
variability, the temperature, the rotational velocity, 
the chemical  abundance, etc.

One possible explanation for the above mentioned differences 
is truncation of the Be star's disk by the compact companion.
The continuous revolution of the neutron star around the Be star primary
prevents the formation of an extended disk in Be/X-ray binaries. This is
expected to occur for short orbital period systems.
As a result of its gravitational field, the compact object not only 
will accrete a part of the disk material, but will distort 
the velocity of the particles at the outer disk. Probably a part of the material
can be pushed in the inner part of the disk. 
This seems quite realistic especially if the neutron star
acts as propeller or ejector, as it was supposed for \lsi\ (Zamanov, 1995).   

For systems with
longer orbital periods ($\simmore \: 100$ days) it seems unlikely that the
neutron star has any influence on the Be star disk. In these systems
the denser disks might be  a selection effect due to the fact that the
neutron star will act as X-ray pulsar  only if the disk is dense. A
neutron star orbiting at long distance around a Be star with low density
disk will act as propeller (accretion onto the magnetosphere) or ejector 
(radio pulsar). If this is the case, a careful  analysis of photospheric
conditions of the long period Be/X-ray stars can  provide us with valuable
information on this issue.

The idea of disk truncation has been put on stronger theoretical grounds by
Negueruela et al. (2000), who interpreted the X-ray and optical
characteristics of the Be/X-ray system 4U 0115+63/V635 Cas as
warping/tilting episodes of the disk. One of the best candidates to
produce the high optically thick disks needed for radiation-driven warping
is disk truncation. 

Finally, while the compact companion does not seem to affect the
formation and initial building-up stages  of the circumstellar disk in
Be/X-ray systems, it does seem to affect its subsequent characteristics
and development. 

\section{Conclusions and open questions}

From our comparative study between the \Ha\ emission lines of Be/X-ray
binaries and those of isolated Be stars we find that:

- The envelops in the Be/X-ray binaries are axisymmetric and 
rotationally supported as in isolated Be stars. 

- The line broadening is similar in the isolated 
Be stars and the Be/X-ray binaries. 

- The circumstellar disks in Be/X-ray binaries are, on average, two times
denser than those in isolated Be stars.

- The relationships between the various \Ha\ parameters (FWHM, $\Delta$V,
EW) seem to indicate the presence of smaller and denser disks in Be/X-ray
binaries than in isolated Be stars, most likely truncated by the compact
object.

Some of the open questions which should be addressed in the future are:

 -- What is the behaviour of the other emission lines in Be/X-ray binaries?
The relative intensities of Balmer lines (Balmer decrement) can give 
more information about the temperature and electron density of the disk. 
The optically thin lines of FeII also can give us important information.

 -- What is the behaviour of the Be binary  systems where the companion is
not a neutron star?

 -- Do the physical properties of the optical companion (metallicity,
 instability in the photosphere induced by binary rotation, etc )
 contribute to producing denser disks? 

\begin{acknowledgements}
We are very grateful to the referee Dr. R. Hanuschik
for the useful comments and suggestions.
~~RZ acknowledges support from Direccion general de relaciones culturales y
cientificas, Spain. JM acknowledge partial support by  
DGESIC (PB97-0903 and PB98-0670-C0201)
and by Junta de Andaluc\'{\i}a (Spain). P. Reig  acknowledges  support 
via  the  European  Union Training and Mobility of Researchers Network
Grant  ERBFMRX/CT98/0195. 

\end{acknowledgements}

\begin{table}
\caption[]{H$\alpha$ line parameters from Southampton -- Valencia data base }

\begin{tabular}{cccccc}
\hline
 Date      &  $-EW($H$\alpha)$ & $\Delta$V     & FWHM         &  Disp.\\
(yyyymmdd) &      (\AA)   & (km s$^{-1}$) &(km s$^{-1}$) & \AA\ pix$^{-1}$ \\ 
\hline
\multicolumn{5}{c}{4U1145-619/V801 Cen}   \\
19850101             &     18.2 &       205 &      363 & 0.8 \\
19850102             &     17.7 &         - &      368 & 0.8 \\
19850103             &     17.6 &         - &      292 & 0.8 \\
19850105             &     16.3 &       205 &        - & 0.8 \\
19930302             &     38.0 &         - &      379 & 0.5 \\
19940307             &     45.0 &        98 &      420 & 0.5 \\
19940702             &     25.0 &         - &      457 & 0.5 \\
19950819             &     27.0 &       216 &      370 & 0.5 \\
19960302             &     16.0 &       174 &      297 & 0.3 \\
19960404             &     14.0 &       227 &      288 & 0.5 \\
[1ex]
\multicolumn{5}{c}{A0535+26/V725 Tau} \\
19880310             &     14.0 &         - &      462 &  2.6 \\
19901114             &      9.9 &       140 &      347 &  1.0 \\
19901227             &      8.6 &       167 &      384 &  0.5 \\
19910828             &      9.4 &       198 &      384 &  0.5 \\
19920818             &      7.4 &       218 &      343 &  0.8 \\
19931206             &     13.0 &         - &      370 &  0.8 \\
19940325             &      9.0 &         - &      343 &  1.3 \\
19940916             &      9.4 &       199 &      361 &  1.3 \\
19941110             &     12.0 &       228 &      393 &  0.8 \\
19950226             &      8.6 &       212 &      361 &  0.5 \\
19950806             &      8.0 &       230 &      402 &  0.4 \\
19951120             &      6.2 &       244 &      416 &  0.5 \\
19960228             &      7.3 &       233 &      416 &  0.4 \\
[1ex]
\multicolumn{5}{c}{V0332+53/BQ Cam} \\
19900128             &      4.5 &         - &      224 &  0.3 \\
19900902             &      5.0 &         - &      293 &  1.0 \\
19901114             &      5.7 &         - &      297 &  1.0 \\
19910127             &      6.5 &         - &      251 &  1.0 \\
19910828             &      4.6 &       143 &      274 &  0.5 \\
19911214             &      4.8 &       153 &      297 &  0.5 \\
19920818             &      3.1 &         - &      265 &  0.8 \\
19921113             &      3.8 &         - &      274 &  0.8 \\
19970814             &      4.2 &         - &      280 &  0.4 \\
19971114             &      3.9 &       155 &      270 &  0.4 \\
[1ex]

\hline
\end{tabular}
\end{table}

\addtocounter{table}{-1}
\begin{table}
\caption[]{Continuation}
\begin{tabular}{cccccc}
\hline
 Date      &  $-EW($H$\alpha)$ & $\Delta$V     & FWHM         & Disp. \\
(yyyymmdd) &      (\AA) & (km s$^{-1}$) &(km s$^{-1}$) & \AA\ pix$^{-1}$  \\ 
[1ex]
\hline

\multicolumn{5}{c}{4U0115+63/V635 Cas} \\
19900214             &      6.8 &       182 &      571 & 1.0 \\
19901227             &      4.5 &       400 &      603 & 0.5 \\
19910127             &      9.5 &         - &      452 & 1.0 \\
19910828             &      3.9 &       453 &      640 & 0.5 \\
19911214             &      5.4 &       497 &      699 & 0.5 \\
19920805             &      4.1 &       425 &      608 & 1.2 \\
19930118             &      7.1 &       406 &      786 & 0.5 \\
19931217             &      5.4 &       362 &      750 & 0.5 \\
19950703             &      3.5 &       448 &      649 & 1.0 \\
19960112             &     11.3 &       215 &      480 & 1.0 \\
19960131             &      8.0 &       292 &      521 & 1.6 \\
19960620             &      7.8 &       321 &      548 & 1.0 \\
19960709             &      6.3 &       404 &      631 & 1.0 \\
19970201             &      3.7 &       470 &        - & 1.0 \\
[1ex]

\multicolumn{5}{c}{A1118-617 / Hen3-640}\\
19930303  & 70.0 &  106 &  384 & 0.5 \\
19930622  & 68.0 &  106 &  407 & 0.6 \\
19940307  & 62.5 &  100 &  398 & 0.5 \\
19940703  & 62.0 &   59 &  416 & 0.5 \\
19950225  & 65.0 &   59 &  384 & 0.4 \\
19960403  & 62.0 &   82 &  398 & 0.5 \\
19970620  & 69.0 &   65 &  384 & 0.4 \\
[1ex]
\multicolumn{5}{c}{RX J0812.4--3114 / LS 992}\\
19980203  & 17.0 &  143 &  457 &  0.4 \\
19990109  & 20.5 &  458 &  457 &  0.4 \\
[1ex]
\multicolumn{5}{c}{4U0728-25} \\
19901227  &  9.0  &   - &  210 & 0.5 \\
19910127  &  7.3  &   - &  293 & 1.0 \\
19921113  &  5.0  & 176 &  361 & 0.8 \\
19930304  &  5.6  &   - &  306 & 0.4 \\
19930624  &  6.0  & 115 &  192 & 0.5 \\
19931205  &  5.6  &   - &  388 & 0.8 \\ 
[1ex] 
\hline
\end{tabular}
\end{table}

\begin{table}
\caption[]{H$\alpha$ line parameters measured on Rozhen spectra}

\begin{tabular}{ccccccc}
\hline
 Date      &  $-EW($H$\alpha)$ & $\Delta$V     & FWHM       & Disp. \\
(yyyymmdd) &    (\AA)        & (km s$^{-1}$) &(km s$^{-1}$) & \AA\ pix$^{-1}$ \\ 
[1ex]
\hline
\multicolumn{5}{c}{XPer / HD 24534} \\
[1ex]
19920903             &       1.9 &       382 &      526 & 0.1 \\
19920905             &       2.2 &       386 &      513 & 0.1 \\
19970817             &      10.5 &        97 &      211 & 0.2 \\
19980209a            &      12.3 &       133 &      287 & 0.2 \\
19980209b            &      12.2 &       134 &      284 & 0.2 \\
19980219a            &      12.1 &       118 &      286 & 0.2 \\
19980219b            &      11.7 &       127 &      286 & 0.2 \\
19981102a            &       9.8 &       142 &      251 & 0.2 \\
19981102b            &       9.9 &       145 &      250 & 0.2 \\
19990309a            &       6.7 &       161 &      283 & 0.2 \\
19990309b            &       6.3 &       164 &      285 & 0.2 \\
19990919a            &       8.1 &       146 &      281 & 0.2 \\
19990919b            &       7.7 &       148 &      282 & 0.2 \\
[1ex]
\multicolumn{5}{c}{\lsi / V615~Cas} \\               
19980930             &     13.3  &  293      &  607     & 0.2 \\
19980930             &     13.4  &  298      &  604     & 0.2 \\  
19981001             &     11.9  &  314      &  600     & 0.2 \\
19981001             &     11.0  &  298      &  617     & 0.2 \\    
19981004             &     12.3  &  346      &  554     & 0.2 \\
19981004             &     12.6  &  345      &  569     & 0.2 \\
19981009             &     10.2  &  330      &  615     & 0.2  \\        
19981209             &      9.3  &  304      &  583     & 0.2  \\
19990105             &      9.3  &  339      &  590     & 0.2  \\       
19990106             &     10.0  &  337      &  585     & 0.2  \\
19990306             &     10.0  &  337      &  629     & 0.2  \\        
19990326             &      9.8  &  327      &  605     & 0.2  \\
19990917             &     10.7  &  368      &  624     & 0.2  \\
19990917             &     10.6  &  353      &  628     & 0.2 \\
19990919             &      9.0  &  347      &  602     & 0.2 \\
19990925             &      9.6  &  340      &  545     & 0.2 \\
19990926             &     10.4  &  327      &  527     & 0.2  \\
19991025             &     12.0  &  319      &  535     & 0.2  \\    
19991126             &     10.7  &  325      &  626     & 0.2  \\
\hline
\end{tabular}
\end{table}

\begin{table}
\caption[]{ Values of $v \sin i$ for the Be/X-ray binaries }
\label{Tvsi}
\begin{tabular}{lcl}
\hline
 Object & \vsi          & Reference    \\ 
        &(km~s$^{-1}$)  &               \\
\hline
               &          &                                    \\ 
    X~Per  \    &   215    &  Lyubimkov et al. (1997)           \\ 
    \lsi\  \    &   360    &  Hutchings \& Crampton (1981a)     \\ 
    \lsa\  \    &   200       &  Reig et al. (1997)     \\  
    4U 1145-619 \ &   270      &  van Paradijs (1995)            \\ 
    A 0535+26   \ &   254      &  Clark et al. (1998)            \\ 
    V 0332+53   \ &   145 $^*$ &  Negueruela et al. (1999)     \\ 
    4U 0115+63  \ &   365      &  Hutchings \& Crampton (1981b)  \\ 
    A 1118-617  \ &   300      &  Janot-Pacheco et al. (1981)    \\ 
    RX J0812-31 \ &   240      &  Reig et al. (2000b)            \\ 
    4U 0728-25  \ &   200      &  Corbet \& Mason  (1984)        \\ 
                \ &            &                                 \\ 

\hline  
\end{tabular}

$\;(^*)$ average value on base of HeI/HeII absorption lines estimates only.

\end{table}

\end{document}